\documentclass[conference]{IEEEtran}
\IEEEoverridecommandlockouts

\usepackage[final]{changes}

\usepackage{balance}

\usepackage{changes}
\usepackage{lipsum}
\definechangesauthor[name={y}, color=orange]{y}
\definechangesauthor[name={b}, color=red]{b}
\usepackage{booktabs}
\newcommand\eat[1]{}

\newcommand{\modelname}{IGNiteR}
\usepackage{multirow}
\usepackage{cite}
\usepackage{bm} 
\usepackage{amsmath,amssymb,amsfonts}
\usepackage{algorithmic}
\usepackage{graphicx}
\usepackage{textcomp}
\usepackage{xcolor}
\usepackage{hyperref}
\usepackage[linesnumbered,ruled,vlined]{algorithm2e}

\SetCommentSty{mycommfont}
\SetKwInput{KwInput}{Input}                
\SetKwInput{KwOutput}{Output}              
\SetKwInput{Parameters}{Parameters}              

\def\BibTeX{{\rm B\kern-.05em{\sc i\kern-.025em b}\kern-.08em
    T\kern-.1667em\lower.7ex\hbox{E}\kern-.125emX}}
\begin{document}

\title{IGNiteR: News Recommendation in Microblogging Applications
}


\author{
\IEEEauthorblockN{Yuting Feng}
\IEEEauthorblockA{\textit{CNRS LISN} \\
\textit{University Paris-Saclay}\\
yuting.feng@universite-paris-saclay.fr}
\and
\IEEEauthorblockN{Bogdan Cautis}
\IEEEauthorblockA{\textit{CNRS LISN and IPAL Singapore} \\
\textit{University Paris-Saclay}\\
bogdan.cautis@universite-paris-saclay.fr}
}

\maketitle

\begin{abstract}
News recommendation is one of the most challenging tasks in recommender systems, mainly due to the ephemeral relevance of news to users. As social media, and particularly microblogging applications like Twitter or Weibo, gains popularity as platforms for news dissemination, personalized news recommendation in this context becomes a significant challenge. We revisit news recommendation in the microblogging scenario, by taking into consideration social interactions and observations tracing how the information that is up for recommendation spreads in an underlying network.  We propose a deep-learning based approach that is diffusion and influence-aware, called \textbf{I}nfluence-\textbf{G}raph \textbf{N}ews \textbf{R}ecommender   (\textbf{I}\textbf{G}\textbf{N}ite\textbf{R}). It is a content-based deep recommendation model that jointly exploits all the data facets that may impact adoption decisions, namely semantics, diffusion-related features pertaining to local and global influence among users, temporal attractiveness, and timeliness, as well as dynamic user preferences. To represent the news, a multi-level attention-based encoder is used to reveal the different interests of users. This news encoder relies on a CNN for the news content and on an attentive LSTM for the diffusion traces. For the latter, by exploiting previously observed news diffusions (cascades) in the microblogging medium, users are mapped to a latent space that captures potential influence on others or susceptibility of being influenced for news adoptions. Similarly, a time-sensitive user encoder enables us to capture the dynamic preferences of users with an attention-based bidirectional LSTM. We perform extensive experiments on two real-world datasets, showing that \modelname\ outperforms the state-of-the-art deep-learning based news recommendation methods.
\end{abstract}


\vspace{-4mm}
\section{Introduction}
\label{sec:intro}
Online recommender systems \cite{ricci2014recommender} have become a ubiquitous tool in our daily lives, allowing us to get the most relevant pieces of information, without the risk of being overloaded by choices. They are also one of the most valuable applications of recent machine learning advances, such as deep learning \cite{salakhutdinov2007restricted,cheng2016wide,guo2017deepfm}. When the items to be recommended are news, additional challenges must be overcome, pertaining to the highly dynamic, ephemeral nature of news. Therefore, specializing recommendation approaches to this specific kind of information has been a beneficial and productive research path, as we have seen in recent years a plethora of techniques for news recommendation, such as \cite{wang2018dkn,zhu2019dan,meng2021dcan,ge2020graph,wu2019npa}.

To further improve recommendation effectiveness, a similar research direction, for methods tailored not necessarily for a specific kind of information, but for a specific scenario of access to information, have been proposed as well, with a recent focus on social recommender systems \cite{fan2019graph,xiao2019beyond}.  In the social media context, it is natural to reason about item relevance by exploiting socially-induced signals. In particular, how the information that is up for recommendation may also spread and reach users in an underlying network through \emph{word-of-mouth} is highly important. The role of word-of-mouth mechanisms – e.g., likes, shares, reposts / retweets, notifications – is twofold: (i) they allow information to propagate easily to a large audience, and (ii) they give credibility to the conveyed messages. Indeed, there are many studies showing that people are more inclined to pay attention to a message or referral coming from a known individual, e.g., a friend or an influencer whose choices he or she may often follow \cite{bughin2015getting}. Accordingly, influence and information diffusion become important dimensions for any recommendation problem in social media. This has generated significant interest in the analysis of information diffusion patterns and influence, under the formal scope of \emph{influence estimation} \cite{DBLP:journals/tkdd/Gomez-RodriguezLK12} and \emph{influence maximization} \cite{DBLP:journals/toc/KempeKT15}. The main applications of information diffusion studies in social networks revolve around the classic user-item ``matching'' problem of recommender system \cite{leskovec2006patterns}, with viral marketing as its most successful example.

In microblogging platforms like Twitter or Weibo, the release, dissemination, and adoption of news items follow a unique pattern, different from the ones of traditional news portals. A news item is firstly posted by a user (the cascade initiator), then it may attract the attention of his / her followers who may in turn repost (adopt) it. This news referral can continue and thus propagate to a large audience. Both the content of the news item and the users involved in its dissemination will determine how far that item may propagate and the extent of its adoption. Intuitively, when a news item reaches a user in the social platform, besides the explicit semantic information it exposes, time has endowed it with other, socially-related information that may sway that user's adoption decision. 

Consequently, as microblogging gains popularity as a platform for news dissemination, \emph{personalized news recommendations} in this scenario becomes an important challenge, one for which the information diffusion patterns and influence mechanisms therein must be well-understood and exploited for effective recommendations. Interestingly, social media is not only the second most important news source (behind TV), accounting for 40\% of the consumption in recent statistics \cite{SNnews} but, in many ways, it has also reshaped the style of news and the users' patterns of news consumption. 

We revisit in this paper the news recommendation problem in the \emph{microblogging} context. In this setting, a specific user may adopt 
a news item not only based on its content / semantics, the personal preferences, or timeliness of that news, but also based on the influence others may exert on her with regards to news adoption. Influence may be exerted either locally (by friends or followed users) or globally (indirectly, by highly influential users). 

In our view, the main limitation of existing ML-based news recommendation approaches, such as \cite{wang2018dkn,zhu2019dan,meng2021dcan,ge2020graph,wu2019npa, wu2019neural,wu2019neural2,wu2020user,an2019neural,wang2020fine,zhang2021unbert}, is that they are generally based on the semantic content of news and on the user profiles, while the underlying recommendation scenario is ignored. We aim to address this limitation and we propose an approach that adopts an influence-aware perspective on news recommendation.  

Our deep-learning based model, called \modelname, requires in the training phase a joint history of news adoptions and news propagation traces (news cascades) in the microblogging application. \modelname\ seeks to exploit jointly all the data facets that may impact news adoption decisions, namely semantics, diffusion-related features pertaining to local and global influence among users, dynamic user preferences, as well as temporal attractiveness and timeliness.

Our main contributions are the following:
\begin{itemize}
    \item We describe how to leverage diffusion cascades to build a behavior-driven user graph and node embeddings for the follow-up recommendation, revealing correlations among users and pinning down an estimate on the probability of information diffusion between them.
    \item We propose to incorporate the influence-level information, based on the participation of users in the news dissemination process, along with semantics, attractiveness, and timeliness for comprehensive news representation.
    \item We design \modelname, in which we 
    attentively fuse the informativeness from different data facets through a news encoder, aggregating the news history with an attention-based sequential model for user profiling.
\end{itemize}
Our experiments with real-world datasets (including a publicly available one) show that  \modelname\ outperforms the state-of-the-art deep-learning based news recommendation methods.

\section{Related Work}
\label{sec:related}
Facing the tremendous volume of online data, a plethora of deep learning-based recommendation models have been developed to deal with problems such as effectiveness, computation cost, sparsity, or scalability. We refer the reader to \cite{zhang2019deep} for an overview on the recent developments in this field, and we focus our related works discussion on the  areas of social-aware or news recommendation techniques.

\subsection{News Recommendation}
\label{sec:news_rec}
News recommendation has been a topic of great interest in the field of recommender systems. The rich textual information of news, their timeliness, diversity, and heterogeneity have all been major challenges for recommendation algorithms, distinguishing them from other recommendation tasks. Nevertheless, the essential problem remains the one of learning a suitable representations of news and users, in order to make accurate recommendations. Benefiting from the recent NLP advances, many state-of-the-art works use pre-trained word embedding representations \cite{le2014distributed,DBLP:conf/naacl/DevlinCLT19} for the textual content of news, addressing the problem of highly condensed semantic information.  In \cite{wu2021empowering}, the authors explore the use of \emph{pre-trained language models} (PLM), in order to mine the deep semantic information of news. Such pre-trained language models are shown to have stronger text modeling ability than shallow models, which are learned from scratch in the
news corpus. Similarly, the work of \cite{zhang2021unbert} learns news  representations by pre-training with the BERT technique \cite{DBLP:conf/naacl/DevlinCLT19}, illustrating the generalization ability to news in cold-start scenarios. 

Besides the semantics dimension, additional information has also been taken into consideration to represent news.  \emph{News categories} are considered in \cite{wu2019npa}'s model, which is further enhanced by a multi-level attention mechanism to fuse information in \cite{wu2019neural},  or by a multi-head self attention mechanism in \cite{wu2019neural2}.  User \emph{dwell time} on news is considered in \cite{wu2020user} . 

Furthermore, external links to \emph{knowledge-level information} have broadened the scope of signals for predicting news adoption. In recent deep-learning based approaches, such diverse signals are extracted and fused via convolutional neural networks (CNNs) \cite{wang2018dkn,zhu2019dan,meng2021dcan}. 
Based on the news representation, users are profiled through the aggregation of (some of) their adopted news, by various methods. For example, in \cite{okura2017embedding}, the authors exploit diverse recurrent networks to model the sequential evolution of user preferences, while \cite{zhu2019dan} uses a Long Short-Term Memory  (LSTM) network to encode clicked news, complemented by an attention layer emphasizing the click history importance. A GRU network is used instead of an LSTM one in \cite{an2019neural}, which proposes to learn and use jointly long-term user representations from the ID embeddings of users, and short-term user representations from the recent browsing history thereof. In \cite{wang2020fine}, the authors propose a fine-grained interest matching method, where each news items in a user's history is endowed with multi-level representations via stacked dilated convolutions.

With the increased focus on Graph Neural Networks (GNN) \cite{DBLP:journals/corr/abs-2003-00982}, as the state-of-the-art approach for analysing and learning on graphs, some recent studies adopt a \emph{graph perspective} for users and news items for recommendation. \cite{ge2020graph} builds a bipartite graph where users and items are  nodes, and the neighbors are aggregated to enhance node representations. In \cite{hu2020graph}, the authors build a users--news--topics graph,  in order to capture the users' long-term interests. In addition to such user--item graphs, works such as \cite{wang2019kgat} incorporate knowledge graphs (KG),  with a graph attention network  modeling the high-order connectivity among users, news, and knowledge entities.

The aforementioned 
models generally ignore the recommendation scenario, being generic by design. Yet by specializing the recommendation approach to specific adoption scenarios, we can gain in effectiveness. In this vein, for news adoptions and recommendation in social platforms, relevant connections between users can be built not only based on commonly clicked news, but also based on social connectivity and, importantly, on the implicit influence exerted among users and the observable diffusion patterns it leads to. To enhance the recommendation process, we place at the core of our model social graph embeddings for users, which can capture local and global influence inferred from information cascades.

\subsection{Social-Aware Recommendation}
\label{sec:social_aware_rec}
The main underlying idea of recommender systems in social media is to capitalize on various social connectivity concepts, such as homophily and influence, and to extract correlations among users.  We discuss next such works, not necessarily pertaining to news. Early recommender system models of this kind leverage the social links as indicators of similarity in collaborative filtering approaches \cite{guo2015trustsvd,ma2011recommender}. Recent deep neural network (DNN) based research focuses on the latent representation of connectivity among users, in order to enhance recommendation performance. For example, \cite{fan2019graph} models social similarity by aggregating a user's social neighbours,  
leading to a social-space latent factor for users. 
The work of \cite{xiao2019beyond} models in a similar way the social space based on each user's social neighbours, with an attention mechanism to aggregate them, with the notable difference that they use a Monte Carlo Tree Search (MCTS) strategy to select the relevant neighborhood. 
In \cite{wu2020diffnet++}, the authors build a GNN to recursively update users and items embeddings, by exploring the social network up to a pre-defined depth, in order to capture higher-order social proximity and influence. Finally, the recent work of \cite{fu2021dual} describes a social-aware recommendation model leveraging the social connections among users to build a user-user graph, complementing the user-item and item-item graphs constructed from the click history.

Beyond social similarity, understanding how information items may be diffused and adopted in sequence by socially connected users (i.e., influence) is of course paramount for effective recommendations in a social media context. Nevertheless, the dissemination paths of news may not strictly follow the known social network topology (e.g., followership in Twitter). Indeed, the reality of social media is much more intertwined, as users may be exposed to information published by others without direct connections, as relationships may span outside the given network structure. In our work, we adopt a deep behavior-driven network to model the social correlations among users, by classifying them into influencers (news post initiators) and influencees (reposters),  according to their parti\-ci\-pation in observed diffusion cascades of microblogging posts about news. We build user node embeddings from a given history of diffusion cascades, as the user representation for the follow-up news recommendation task.

\section{Problem Formulation}
\label{sec:problem formulation}

Generally, in news recommendation, a given user $i$ has an adoption history consisting of a set or a sequence of news items $x_{1},x_{2},...,x_{s_{i}}$, 
and the recommendation task is to predict the probability that $i$ will adopt (e.g., click on) some unseen candidate news $x$. In the microblogging context, the notion of click is replaced by the one of posting / tweeting. In such a practical context, a piece of news is first posted by the diffusion (cascade) initiator, and then adopted (posted) by other users involved in that diffusion process. When a cascade reaches a user, he or she will be notified that friends / followees adopted that news item, increasing  awareness about it and thus influencing the adoption decision.

In our study, the overall input for our news recommendation framework consists of (i) a follower  graph $G=(V,E)$, where $V$ are the nodes (users) and $E$ are the followship edges, possibly enriched by various node or edge features, and (ii) a cascade history $\mathcal{H}$, i.e., a set of diffusion cascades, which gives us the timed news adoptions, possibly enriched by various news features. A diffusion cascade for item $x$ is a time-ordered sequence of adoptions
\begin{equation}
    C_x = \left[(v_0, t_0), (v_1,t_1),\dots(v_{m^x}, t_{m^x})\right]
\end{equation}
 initiated by influencer $v_0$ at time $t_0$, with all $v_j$, $j\geq 1$, being the reposters and  $m^x$ denoting the number of reposters of $x$. From the cascades $\mathcal{H}$, we also distinguish the overall subsets of users $V_{influencer} \subseteq V$ and  $V_{reposter} \subseteq V$.


When a target user $i$ is exposed to a news item $x$ at time $t$,  several data facets of $x$ may contribute to $i$'s adoption decision: (i) the news content, (ii) the users already involved in the diffusion process of $x$ up to moment $t$, (iii) the attractiveness of $x$,  and (iv) the timeliness (lifespan) of $x$.  More formally, a news item  $x$ at time $t$ has as raw (initial) representation $\phi^{x}_t$, which will be short notation for the tuple $(s^x,V^x_{t},A^x_{t})$, where $s^x$ is the semantic description of $x$,  the sequence $V^x_{t} = \left[v_0,v_1,\dots,v_{m_t^x}\right]$ represents the users who participated in the diffusion process of $x$ up to moment $t$\footnote{A projection of $C_x$ on the user dimension, up to moment $t$.}, with $v_0$ the initial user who published the news (initiator of the propagation chain), and $A_t$ represents attractiveness (captured in our training data layout by  histograms on the number of adoptions over time). 

Therefore, the recommendation task becomes one of predicting whether a target user $i$ will adopt at time $t$ the candidate news $x$, described initially by $\phi^{x}_{t}$, based on $i$'s adoption history 
$\{\phi_{t_1}^{x_1},\phi_{t_2}^{x_2}, \dots,\phi_{t_{s_i}}^{x_{s_i}}\}$, where  each
$\phi_{t_j}^{x_j}$ represents the state of a news $x_j$ when it was adopted by  $i$ at time $t_j$, $t_j < t$, with $s_i$ being the length of $i$'s relevant history.

\section{Our Approach}
\label{sec:model}

\begin{figure*}[t!]
\vspace{-6mm}
  \centering
  \includegraphics[width=0.8\linewidth]{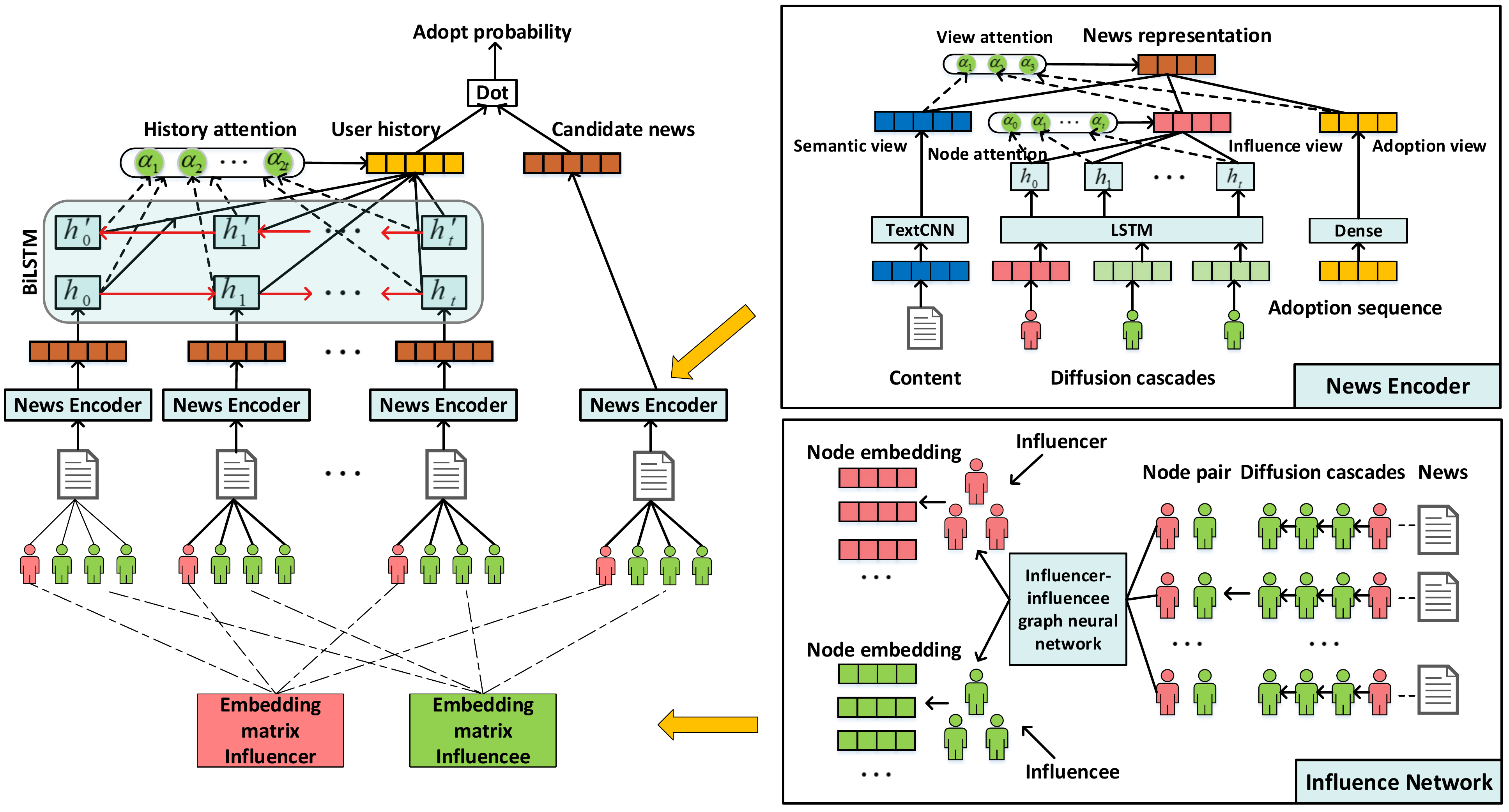}
  \vspace{-2mm}
  \caption{The framework of IGNiteR.}
  \label{fig:framework}
\vspace{-4mm}
\end{figure*}

We describe in this section our approach for influence-aware news recommendation, starting with the latent representation of nodes extracted from news cascades (Sec. \ref{sec:user graph embedding}). In Sec. \ref{sec:local influence}, we describe how we focus the diffusion of a news item $x$ to a view that is personalized to the target user. In Sec. \ref{sec:newsEncoder} we present the multi-level attention mechanism for encoding the news, and in Sec. \ref{sec:userEncoder} we complete our framework with the time-sensitive user encoding aspects. Finally, we conclude this section with the discussion on other notable training details (Sec. \ref{sec:training}). Fig. \ref{fig:framework} illustrates the overall framework.

\subsection{{\modelname} Influence-Aware User Graph Neural Network}
\label{sec:user graph embedding}
In the various recommendation works that rely on GNNs applied to user graphs or user-item graphs, the users usually have a latent representation that captures their interests \cite{wu2019npa,ge2020graph,wang2019neural}, by a trainable embedding matrix representing each user as a low-dimensional vector. While representations capture mostly the social-connectivity induced features, they fail to encompass aspects pertaining to how information is diffused and how influence,  whether local or global, may be exerted between users. When news cascades can be observed and exploited, higher-order connections and latent representations can be inferred, which are closely tied to news adoptions.

\begin{figure}[t!]
\vspace{-3mm}
  \centering
  \includegraphics[width=0.9\linewidth]{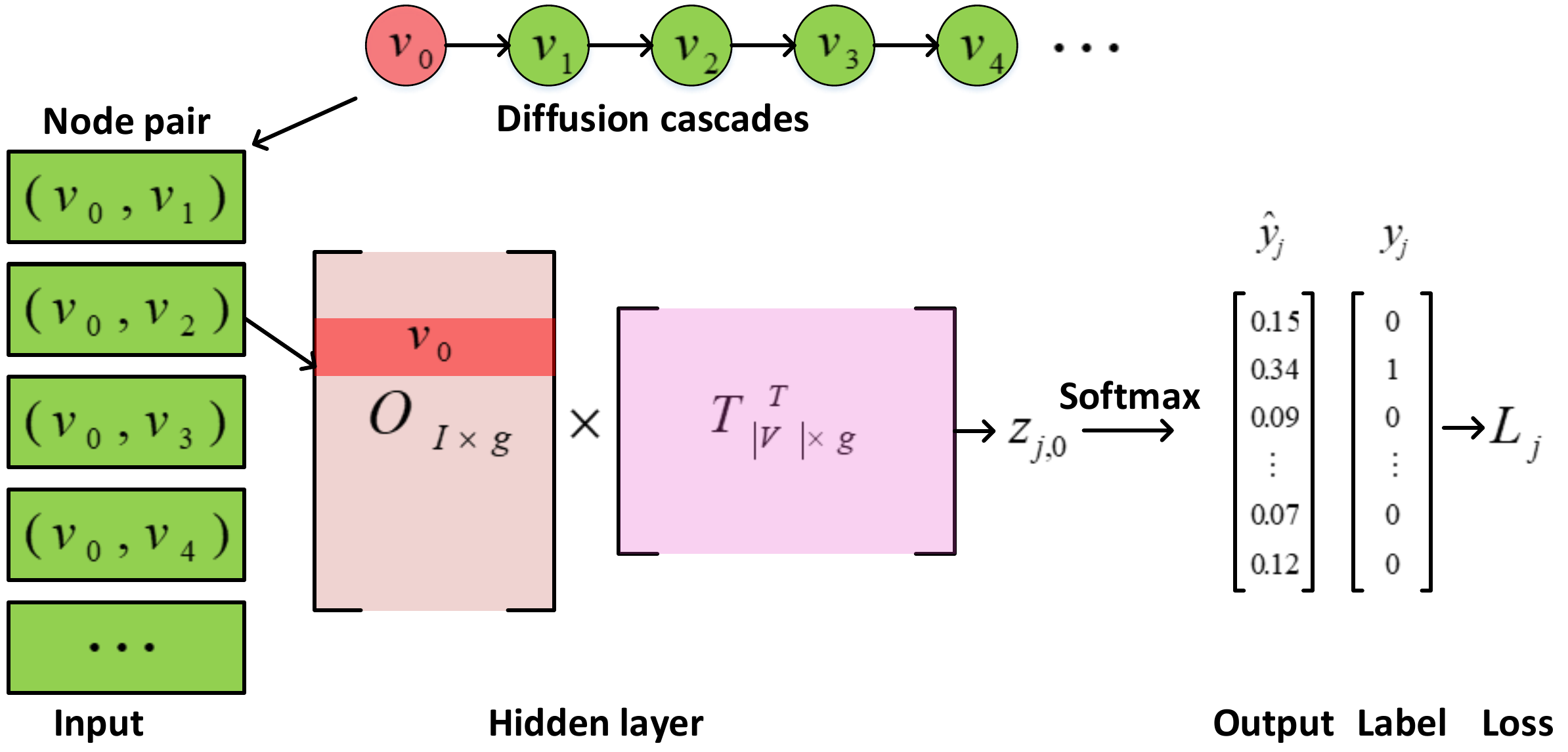}
  \vspace{-2mm}
  \caption{Influence-graph representation from news cascades.}
\vspace{-5mm}
\label{fig: userGNN}
\end{figure}

Inspired by the recent work of  \cite{panagopoulos2020multi}, on influence maximization with node representations learned from cascades, we train a neural network in order to obtain embeddings for influencers (cascade initiators) and for reposters of news.  While the goal of influence maximization is to find the set of influencers maximizing the spread of information in a diffusion network, our goal -- diffusion-aware news recommendation in the microblogging scenario -- can rely on similar means,   i.e, diffusion-aware latent node representations.

Information about an influencer's aptitude in swaying the adoption decision of others can be derived from historical cascades.  The original context is extracted from cascades, where the input feature is the cascade initiator and the output labels are the set of participating nodes in that cascade (Fig. \ref{fig: userGNN}). We down-sample this context, based on the temporal information of actions taken by the cascade participants, as  rapidly responding participants are more likely to be influenced. More precisely, from a given cascade $C_x$ of news item $x$,
\begin{equation*}
    \left[(v_0,t_0),\dots,(v_i,t_i),\dots,(v_{m_x}, t_{m_x})\right],
\end{equation*}
the probability of sampling a reposter  $v_j$ is inversely proportional to the elapsed time $t_0 - t_j$ since $C_x$ was initiated: 
\begin{align}\label{eq:5}
    p\left(v_j|x\right) = \frac{\left(t_j - t_0\right)^{-1}}{\sum_{i}\left(t_i - t_0\right)^{-1}}.
\end{align}

All the nodes are initiated with one-hot embedding, and $\bm{O}\in \mathbb{R}^{I\times g}$ and $\bm{T}\in \mathbb{R}^{|V|\times g}$ are respectively the embedding matrix of cascade initiators and reposters, with $|V|$ the number of nodes in the network,  $I=|V_{influencer}|$, and $g$ being the embedding dimension. With the hidden layer output $\bm{z}_{j,0} = \bm{O}_0\bm{T}^{T} + \bm{b}_j$, where $\bm{b}_t$ denotes the bias, the output of the model is the overall (i.e., multi-hop) diffusion probability $P_{0,j}$ that node $v_j$ may be influenced by the influencer $v_0$. Through a softmax function, the output of the model becomes
\begin{equation}
    \hat{\bm{y}_j}=\frac{e^{\bm{z}_{j,0}}}{\sum_{v_{0'}\in V}e^{\bm{z}_{j,0'}}}
\end{equation}
with $\bm{y}_j$ being the one-hot representation of target node $v_j$, in order to minimize the cross-entropy loss function
\begin{equation}
    L_j = \bm{y}_j\mathrm{log}(\hat{\bm{y}_j}).
\end{equation}
The trained matrices $\bm{O}$ and $\bm{T}$ give us separately the node embeddings of influencers and reposters in the social network, which will then be used for the subsequent recommendation task, with users belonging to either $V_{\text{influencer}}$ or $V_{\text{reposter}}$.

\subsection{Personalized Cascade Views}
\label{sec:local influence}
As stated in Sec. \ref{sec:problem formulation}, the initial description $\phi^{x}_{t_0}$ of a news item $x$ at its publication time $t_0$ corresponds to a tuple $(s^x,V_{t_0}^x,A_{t_0}^x)$, where $V_{t_0}^x = \left[v_0\right]$. By spreading  in the social network, by what can be seen as a snowball effect, the news item carries richer and richer information as more users get involved, such that its  raw representation $\phi^{x}_{t}$  at time $t$ becomes $(s^x,V_{t}^x,A_{t}^x)$, with $V_{t}^x = \left[v_0,v_1,\dots \right]$ 
now a potentially long  time-ordered  sequence of nodes. 

However, for the perspective of the target user $i$, not every user in this sequence may  necessarily exert a significant influence on $i$. With this in mind, and in order to also limit the computation cost of training on long diffusion chains,  we select for user $i$ a most representative sub-sequence $V_{t}^{i}$ of fixed size $m$ (one of the parameters of our model), from the initial cascade sequence $V_t$. In short, we keep the most influential nodes and the closest neighbors of $i$, either by connectivity in $G$ or by similarity score from the reposter embedding matrix $T$.   The detailed sampling procedure for $i$'s perspective on an incoming cascade is given in Alg. \ref{alg:node sampling}. In matrix $T$, the vector  $\bm{v_i}$ represents the $i$-th row, i.e., reposter embedding of user $i$.

\begin{algorithm}[t]
\caption{Local influence selection}\label{alg:node sampling}
  \KwInput{Sequence $V_t$, $G$=$(V,E)$, matrix $\bm{T}$,  $V_{\text{influencer}}$,  node $i$}
  \KwOutput{New cascade sequence $V_{t}^{i}$}
  \Parameters{$m$}

$V_{t}^{i} \gets \left[\right]$\\
\For{$v_j\in V_t$}
{\tcc{keep  influencers and social neighbors}
\If{$v_{j}\in V_{\text{influencer}}$ $\textbf{or}$ $e_{ij} \in E$}
{
$V_{t}^{i}\text{.append}(v_j)$\;
$V_{t}.\text{delete}(v_j)$
}
}
\If{$|V_{t}^{i}|< m$}
{
\tcc{add nearest neighbors of nodes  in $V_{t}^{i}$}
$L\gets \left[\right]$\;
\For{$v_j\in V_t$}
{
$\text{sim} \gets \bm{v_i}^{T}\bm{v_j}$\;
$L\text{.append}((v_j,sim))$
}
$L\gets \text{sort}(L,1)$ \tcp*{sort by similarity}
$L\gets L\left[0:(m-|V_{t}^{i}|)\right]$\;
$V_{t}^{i}\text{.extend}(L\left[:,0\right])$\;
}
\Return $V_{t}^{i}$
\end{algorithm}

With the refined node sequence $V^{i}_{t} =  \left[v_0, v^{'}_1,\dots,v^{'}_{m}\right]$\footnote{Here, the superscript $i$ replaces the $x$ one, to denote user $i$'s perspective on the cascade chain of news item $x$.}, $v^{'}_j\in V_t$, the raw description $\phi^{x}_t$ of news $x$ will thus be given by the tuple $(s^x,V^{i}_{t},A_{t}^x)$. To simplify notation,  whenever  user $i$ and news $x$ are implicitly assumed in the following, the diffusion sequence will simply be denoted by $V_{t} = \left[v_0,v_1,\dots,v_{m}\right]$ and the news information $\phi_t$ by $(s,V_{t},A_{t})$.

\subsection{Multi-Level Attention-based News Encoder}
\label{sec:newsEncoder}
As described previously, a candidate news item's state  $\phi_t$  includes  information from multiple channels, in order to capture not only semantics, but also the diffusion history and temporal information, via the propagation chain ($V_t$) and the distribution of adoptions over time ($A_t$). 
Accordingly, we designed a \emph{time-sensitive news encoder} to comprehensively encode the news items, starting from the raw description $\phi_t$.

\subsubsection{Semantic Information Extraction}
For the semantic facet of news,  we use a one-dimensional CNN \cite{kim-2014-convolutional} to extract the semantic information from text. We focus here on the news title (this can be easily extended to  the abstract / article)  seen as a sequence of words $s=\{w_1,w_2,\dots,w_{n}\}$, with $n$ denoting the title length. We use a pre-trained word embedding method, based on a  large corpus, in order to get an overall word-embedding matrix $\bm{S} = \left[\bm{w}_1,\bm{w}_2,\dots,\bm{w}_{n}\right]^T \in \mathbb{R}^{n \times g_1}$, where $g_1$ is the word-embedding dimension.

The CNN uses multiple filters on the word-embedding ma\-trix $\bm{S}$, each filter $\bm{h}\in\mathbb{R}^{g_1\times{l}}$ being applied on the sub-matrix $\bm{S}_{i+l-1}$, with varying window size $l$, to get a new feature
\begin{equation}
    c_i^h = f(\bm{h}\cdot\bm{S}_{i:i+l-1}+\bm{b})
\end{equation}
with $i=1,2, \dots,n-l+1$ and the bias $\bm{b}\in\mathbb{R}^g_1$. As the filter goes in the direction of sentence length, a feature map $[c_1^h,c_2^h,\cdots,c_{n-l+1}^h]$ is obtained for each filter; then, by max-pooling, we can choose the most representative feature
\begin{equation}
    c_{max}^h = \mathrm{max}\{c_1^h,c_2^h,\cdots,c_{n-l+1}^h\}.
\end{equation}
The final encoding of news semantics  is given by the concatenation of the max-pooling results of $\gamma$ filters, denoted as
\begin{equation}
    \bm{e^s} = [c_{max}^{h_1},c_{max}^{h_2},\cdots,c_{max}^{h_\gamma}].
\end{equation}

\subsubsection{Diffusion Cascade Aggregation}
\label{sec:news nodes}
Recall that another fa\-cet in a news's raw representation comes to its  cascade.  As discussed in Sec. \ref{sec:local influence}, we first distill in the diffusion chain the most representative nodes related to the target user. 

Then, we aggregate the node sequence of the resulting fixed-size cascade $V_t$ into a vector space that can be paralleled with the semantic representation of the news item. For  this stage, we use an LSTM model as the encoder of $V_{t}$, since it is a sequence of time-ordered users, thus capturing temporal correlations and dependencies among users. In addition, as the influence that each node in $V_t$ may exert on the target user $i$ may vary considerably, we add an attention network on the output of hidden layer, to build an attention-based LSTM model that can emphasize potential variations in input relevance.

Given the node sequence $V_{t} = \left[v_0,v_1,\dots,v_j,\dots,v_{m}\right]$, the node embedding sequence (as explained in Sec. \ref{sec:user graph embedding}) can be written as $\bm{V} = \left[\bm{v}_0,\bm{v}_1,...\bm{v}_{m}\right]^T \in \mathbb{R}^{m\times{g_2}}$, where $g_2$ denotes the node embedding dimension. Considering the node sequence as an input of $m$ time steps, each cell in the LSTM can be computed as follows:
\begin{flalign}
\bm{f}_t &= \sigma(\bm{W}_f\left[\bm{h}_{t-1},\bm{v}_t\right] + \bm{b}_f)\\
\bm{i}_t &= \sigma(\bm{W}_i\left[\bm{h}_{t-1},\bm{v}_t\right] + \bm{b}_i)\\
\bm{o}_t &= \sigma(\bm{W}_o\left[\bm{h}_{t-1},\bm{v}_t\right] + \bm{b}_o)\\
\widetilde {\bm{C}}_t &= \tanh{(\bm{W}_C\left[\bm{h}_{t-1},\bm{v}_t\right] + \bm{b}_C)}\\
\bm{C}_t &= \bm{f}_t \odot \bm{C}_{t-1} + \bm{i}_t \odot \widetilde {\bm{C}}_t\\
\bm{h}_t &= \bm{o}_t \odot \tanh{\bm{C}_t}
\end{flalign}
where $\bm{h}\in \mathbb{R}^u$ is the hidden state, $\bm{W}_f$, $\bm{W}_i$, $\bm{W}_o$, $\bm{W}_C\in \mathbb{R}^{(u+g_2)\times u}$ are the weighted matrices and $\bm{b}_f$, $\bm{b}_i$, $\bm{b}_o$, $\bm{b}_C\in \mathbb{R}^{u}$ are the biases of the LSTM, trained to parameterize the forget, input, output gates, and block input respectively. $u$ is the number of units in an LSTM cell.

Instead of getting the final step, we retain the hidden state of each time-step into a sequence $\left[\bm{h}_0,\bm{h}_1,\cdots,\bm{h}_m\right]$, and apply an attention mechanism to aggregate the output of LSTM cells:
\begin{flalign}
h_i &= {\bm{q}_h}^T\tanh(\bm{W}_h\bm{h}_i + \bm{b}_h)\\
\alpha_{i}^h &= \mathrm{softmax}(h_i)= \frac{\mathrm{exp}(h_i)}{\sum_{j=0}^{m}\mathrm{exp}(h_j)}
\end{flalign}
where $\bm{W}_h\in \mathbb{R}^{u\times u}$, $\bm{b}_h\in \mathbb{R}^{u}$ are the projection parameters, and $\bm{q}_h\in \mathbb{R}^{u}$ is the attention query vector. With the attention weights, the node sequence is expressed as the weighted representation of the hidden state from each time step 
\begin{equation}
    \bm{e}^v=\sum_{j=0}^{m}\alpha_{i}^h\bm{h}_j.
\end{equation}

\subsubsection{News Adoption Sequence and Lifespan}
\begin{figure}[t!]
\vspace{-3mm}
  \centering
  \includegraphics[width=0.8\linewidth]{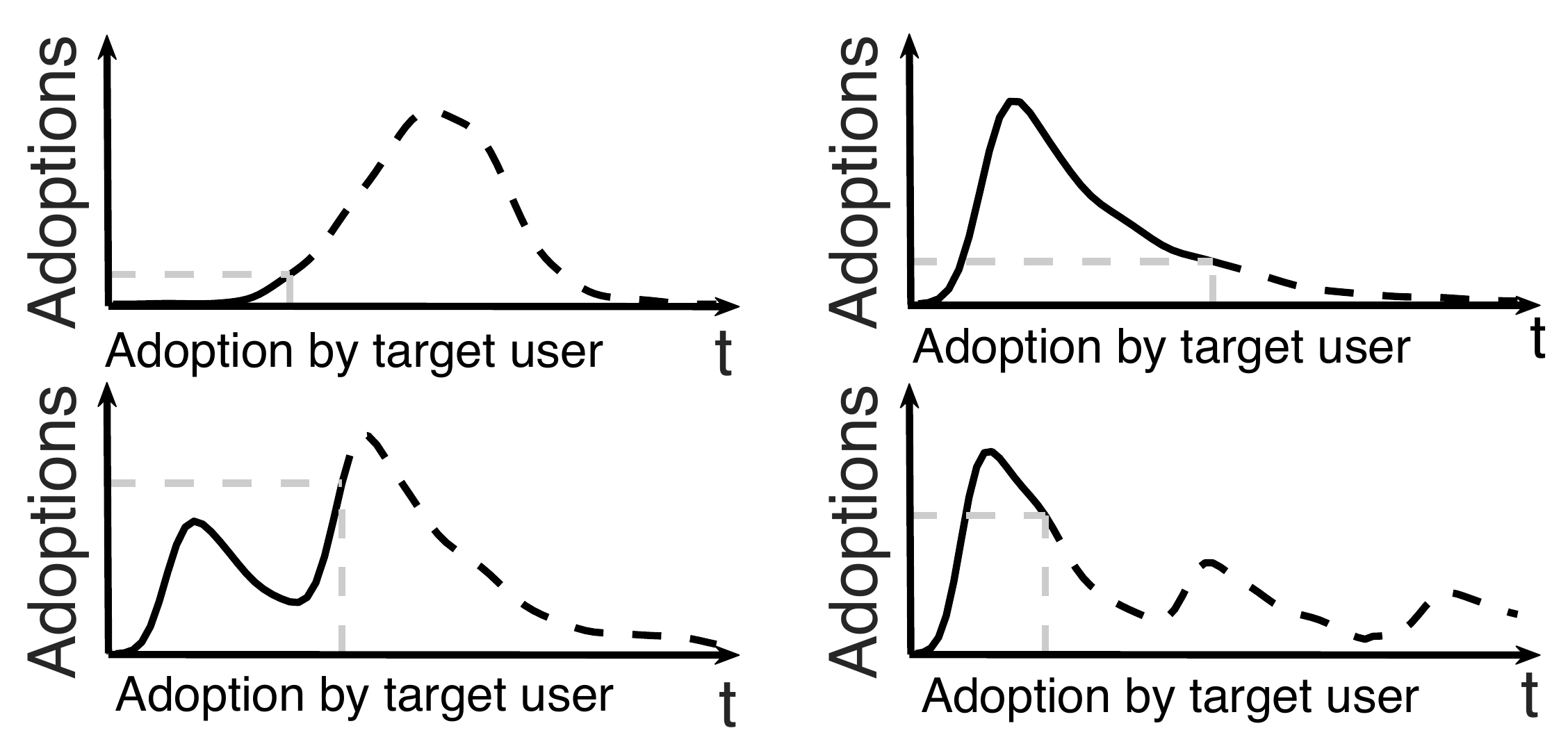}
  \vspace{-2mm}
  \caption{User adoption at various stages in the evolution chart of news items.}
  \label{fig:adoption chart}
\vspace{-3mm}
\end{figure}

Besides the semantics and diffusion dimensions of news, we need to account for  attractiveness and timeliness, as undoubtedly people tend to be attracted by ``hot'' stories. The distribution of the number of adoptions over time can be used as an indicator of popularity and evolution trend -- what we call \emph{attractiveness}, while the timeliness of news can be captured by the lifespan since the initial posting. Moreover, as the adoption chart traces the adoption evolution pattern of a news item, differences in the target user's reaction to such a news item at a specific moment in the charted evolution may be indicative of different news types / categories  and preferences thereof \cite{castillo2014characterizing}. Fig. \ref{fig:adoption chart} exhibits such different situations (this is for illustration only, it does not correspond to a specific user in our datasets).

We use the sequence $A_t=\left[(a_0, {t_0}),(a_1, {t_1}),\dots,(a_d,{t_d})\right]$ of adoptions, where $a_i$ records the number of adoptions (retweets) during a time step, and $t_i$ is the corresponding timestamp. 
Considering that the number of adoptions can vary significantly and may be very large in certain cases, and that news popularity decays with time, we calibrate the raw counts $a_i$ as $a_i^{'} = {f(a_i)}{(t_d-t_0)}^{-1}$, where $f$ is the scaling function, $t_d$ is the current moment, and $t_0$ is the publication time of the item. The adoption vector becomes $e^a = \left[a^{'}_1,a^{'}_2,\dots,a^{'}_d\right]$. 

\subsubsection{Multi-Views Attentive Fusion}
\label{sec:multi-view attention}
In the end, we propose a macro-view attention network to consolidate the three components from different sources. Firstly, we align the dimensions of the three vectors $\left[e^s,e^v,e^a\right]$ to the same vector space $\mathbb{R}^g$, through linear / non-linear transformations for $e^s$ and $e^v$, while padding is used for the adoption sequence $e^a$. Then, we build an attention network to allocate weights to the semantics information, influence representation, and adoption evolution pattern respectively, noted as $\alpha_s$, $\alpha_v$, and $\alpha_a$. For illustration, the semantics weight is computed with:
\begin{flalign}
a_s &= {\bm{q}_s}^T\tanh(\bm{W}_s\bm{e}_s + \bm{b}_s)\\
\alpha_{s} &= \frac{\mathrm{exp}(a_s)}{\mathrm{exp}(a_s) + \mathrm{exp}(a_v) + \mathrm{exp}(a_a)},
\end{flalign}
where  $\bm{W}_s\in \mathbb{R}^{g\times g}$ and $\bm{b}_h\in \mathbb{R}^{g}$ are the projection parameters, while $\bm{q}_s\in \mathbb{R}^{g}$ is the attention query vector. The attention weights for other views can be obtained in a similar way, and the final weighted encoding of a news item is as follows:
\begin{equation}
    \bm{e}^{n} = \alpha_{s}\bm{e}^{s}+ \alpha_{v}\bm{e}^{v} + \alpha_{a}\bm{e}^{a} 
\end{equation}
\subsection{Time-Sensitive User Encoder}
\label{sec:userEncoder}
In accordance with the problem formulation, for a target user $i$ with an adoption history 
$\{\phi_{t_1}^{x_1},\phi_{t_2}^{x_2}, \dots,\phi_{t_{s_i}}^{x_{s_i}}\}$, the multi-level attention news encoder is applied to each piece of news, such that the adoption history becomes $\{e^{1},e^{2},...,e^{s}\}$ (in order to avoid cluttering, the $i$ subscript is omitted in what follows). Following the methodology from Sec. \ref{sec:news nodes}, we build an analogous model to encode users.

The adoption history is a time-ordered sequence of news items, illustrating the evolution of  user's preferences over time. Unlike a diffusion cascade, where the propagation is uni-directional, a user may still be interested in a topic that appears early in the history.  To represent the changes in interests and temporal dependencies among news, we adopt a bidirectional LSTM (BiLSTM) to use the ``past'' for ``future'' in the forward phase and vice-versa in the backward phase of the training.

The BiLSTM mechanism is similar to the LSTM, except that the hidden state output is doubled, so information from both ends is preserved.  We thus obtain the hidden state sequence as $\left[\bm{h}_0,\bm{h}_1,\cdots,\bm{h}_{2s}\right]$. 

Given the adoption history for the target user, relevance of different items therein to the candidate news item may va\-ry, i.e., different items may have a different impact on the adoption decision. To deal with the potentially diverse range of interests of users, an attention network is added upon the hidden state of the BiLSTM,  to get the final weighted encoding
\begin{flalign}
h_{i}^{B} &= {\bm{q}_B}^T\tanh(\bm{W}_B\bm{h}_i^{B} + \bm{b}_B)\\
\alpha_{i}^{h^{B}} &= \mathrm{softmax}(h_{i}^{B})= \frac{\mathrm{exp}(h_i)}{\sum_{j=1}^{2s}\mathrm{exp}(h_j^{B})},
\end{flalign}
where  $\bm{W}_B\in \mathbb{R}^{2s\times 2s}$ and $\bm{b}_B\in \mathbb{R}^{2s}$ are the projection parameters, while $\bm{q}_B\in \mathbb{R}^{2s}$ is the attention query vector. The final user encoding becomes the following:
\begin{equation}
    \bm{e}^u=\sum_{j=1}^{2s}\alpha_{i}^h\bm{h}_{j}^{B}.
\end{equation}

\subsection{Model Training}
\label{sec:training}
In a social networking environment, users may be overwhelmed by a plethora of information received by various notification mechanisms. Indeed, in order to increase the diversity of news recommendations and maximize adoption likelihood, notifications for many news stories may be presented in an impression. This leads to a common phenomenon that users may click only a single or a few pieces of news among the displayed ones. In order to cope with this bias, some neural news recommendation approaches \cite{wang2018dkn,zhu2019dan} manually balance the positive and negative samples. 
However, the informativeness of negative news samples 
should be taken into account.  Here, we follow a negative sampling strategy similar to that of  \cite{wu2019neural,zhai2016deepintent,ge2020graph}, in order to simulate the real-word  scenarios for news exposure and adoption.

In the training phase, we generate $1+\lambda$ news items as an impression, among which one  item is sampled from the user's history as a positive sample, and the remaining $\lambda$ items are negative samples. Then, the recommendation task is reformulated as a $1+\lambda$ multi-class classification. With the softmax function to normalize the adoption probability for each ``class'', the final adoption probability on the positive sample is expressed as follows:
\begin{equation}
    p_i = \frac{\mathrm{exp}(\hat{y}^+)}{\mathrm{exp}(\hat{y}^+) + \sum_{i=1}^{\lambda}\mathrm{exp}(\hat{y}_{i}^-)},
\end{equation}
where $\hat{y}$ is the inner product between the sample news representation and the user representation, while $+$ denotes the positive sample and $-$ denotes the negative ones in the session. The loss function in the classification-like training becomes the minimization of the log-likelihood on all the positive samples:
\begin{equation}
    \mathcal{L} = - \sum_{i=1}^{s}\log(p_i),
\end{equation}
$s$ being the number of positive samples, i.e.,  the history length. 

\vspace{-2mm}
\section{Experiments}
\vspace{-1mm}
We use datasets from the two main microblogging platforms,  Twitter and Weibo, where users post and interact with messages that we will generically call ``tweets''. We collected the  Twitter dataset through its API, while the Weibo dataset is a publicly available one \cite{zhang2013social}. Since news appear implicitly in tweets as links, pointing to the original publisher page, we first identify these links in tweets, and then we crawl the news articles from the corresponding pages. 

Microblogging  provides  diffusion-oriented information, consisting of social activities (tweet / retweet behaviour) pertaining to news, and we can trace the diffusion path and adoption trend during a news item's lifespan, by observing the involved users. The main statistics for the two datasets, following some pre-processing and filtering steps detailed in Sec. \ref{sec:exp_setting}, are presented in Table \ref{tab:tab_datasets}. 

For reproducibility, the  \modelname\ code (including the entire data processing pipeline, from the raw data to the experimental results) and the Weibo data are available at the following anonymous repository: \url{https://github.com/goldenretriever-5423/IGNiteR}. 

\begin{table}
\vspace{-3mm}
  \caption{Dataset Description}
  \label{tab:tab_datasets}
  \begin{tabular}{lll}
    \toprule
         & {\itshape Twitter} &{\itshape Weibo} \\
    \midrule
          number of users & 248,195 & 692,833\\
          number of retweet records & 4,999,535 & 31,211,347 \\
         number of original tweets & 4,566,942& 232,978 \\
          number of news &441,632&13,153 \\
         average number of words per title & 6.94 & 7.26\\
         median length of diffusion chain & 2.74 & 23\\
         maximal length of diffusion chain & 124 & 31009\\
  \bottomrule
\end{tabular}
\vspace{-3mm}
\end{table}

\vspace{-1mm}
\subsection{\added[id=y]{Comparison Models}}
\label{sec:baselines}
We compare with the following state-of-the-art methods, thoroughly fine-tuned in order to obtain their best performance.
\subsubsection{\added[id=y]{Generic recommendation models}}
\textbf{LibFM} \cite{rendle2012factorization} is a classical factorization model used in feature engineering to estimate interactions. Here, news title embeddings and node embeddings of users are used as features. 
\textbf{DeepWide} \cite{cheng2016wide} is a deep learning model using a non-linear part (deep) and a linear one (wide) to learn feature interactions. News titles, user nodes, and concatenated adoption sequences are its input. 
\textbf{DeepFM} \cite{guo2017deepfm} is an end-to-end factorization machine-based deep neural model, using a shared input with ``wide'' and ``deep'' parts similar in spirit to DeepWide; it uses the same input as \textbf{DeepWide}. Also having the same input as \textbf{DeepWide}, 
\textbf{DCN} \cite{wang2020dcn} uses a  cross network for learning bounded-degree feature interactions, while maintaining the benefit of DNNs on high dimensional non-linear features. 
\subsubsection{\added[id=y]{Deep neural news recommendation models}}
\textbf{DKN} \cite{wang2018dkn} recommends news by exploiting a  knowledge graph to capture relationships between news, and links the candidate news with the target user's history of adopted news both at knowledge-level and at semantic-level. News titles and knowledge entities therein are used for this method. \textbf{DAN} \cite{zhu2019dan} enhances \textbf{DKN}'s framework with knowledge-entity information and an LSTM mechanism to model the sequential evolution of user interests. Accordingly, for this method, we use knowledge entities, entity types, and an LSTM  to aggregate the users' history. 
\textbf{GERL} \cite{ge2020graph} builds a bipartite graph  where users and news are  nodes, and neighboring nodes are respectively aggregated in the representation of news or users, along with the semantic information for the recommendation. For this method,  we build the same kind of graph. 
\textbf{NAML} \cite{wu2019neural} exploits semantics and news categories, by a multi-level attention mechanism to fuse information. We use the topic distribution vector for categories and the news' titles for semantics. 
\cite{wu2021empowering} \added[id=y]{revisits the text modeling of news by pre-trained language models (PLMs), and we consider here  the \textbf{NAML-BERT} variant described in \cite{wu2021empowering}.}  \textbf{LSTUR} \cite{an2019neural} \added[id=y]{learns long-term user representations from ID embeddings of users, and short-term user representations from their recently browsed news, by a GRU network. } 
\textbf{FIM} \cite{wang2020fine} \added[id=y]{is a fine-grained interest matching method, where each news items in the user history is endowed with multi-level representations,  via stacked dilated convolutions.}

\subsubsection{\added[id=y]{Social-aware recommendation model}}
\added[id=y]{While the other news recommendation methods exploit mainly the informativeness of the semantics dimension, given the news recommendation scenario we consider, we also select as a baseline method one that exploits the social dimension of news adoptions.} We use the \textbf{DICER} approach of \cite{fu2021dual}, \added[id=y]{a state-of-the-art social-aware recommendation model, which leverages the social connections among users to build a user-user graph, complementing the user-item and item-item graphs constructed from the click history. To compensate for the semantics dimension, we replace in this model the ID embedding matrix of news with the averaged news title embeddings.}

\subsubsection{\added[id=y]{Ablation variants}}
\added[id=y]{To verify the effectiveness of two key dimensions of our model, namely diffusion  and attractiveness, we also describe an ablation study removing from the news' raw representation either the diffusion  sequence $V_t$ (\textbf{{\modelname}-$\mathbf{V_{t}}$}) or the adoption sequence $A_t$ (\textbf{{\modelname}-$\mathbf{A_{t}}$}).}

\added[id=y]{Note: we did not compare with the NRMS model of \cite{wu2019neural2},  as it bears significant similarities with  NAML on the attention mechanism. Also, we did not have access to an open source code for the UNBERT  model of \cite{zhang2021unbert}, which conceptually is similar to  NAML-BERT.}

\subsection{Experimental Setting}
 \label{sec:exp_setting}
\modelname\ was implemented using Tensorflow. \deleted{From the initial data, we keep only users with at least $20$ adopted news.} Since the two datasets record the users' behaviours over 3 years, we \deleted{Then, for each user and each training sample, we} observe data within 3-months windows size, and in this way, a user may be split into several users. \added[id=y]{We split the train / test data by the timeline, so that around 85\% data is used for training and 15\% for testing, while 10\% from the train set is used for validation.} We randomly sampled news released during this period and not posted by the target user as negative samples. For each impression, the ratio $\lambda$ is set to $4$, i.e., we use one positive sample and $4$ negative ones for a training impression. To generate personalized cascade views, we set the count for representative nodes to $m=30$. To avoid outliers, we also apply other filters as follows: the maximal number of adopted news items per user \added[id=y]{in the selected window size} is set to $20$, the maximum title length is set to $20$, the adoption time unit is set to one hour, the maximum adoption length is set to $120$, and the maximum number of negative samples in testing phase is set to $10$. Considering the highly skewed adoption patterns for popular news, we use \emph{two-times log} as the scaling function $f$. 
 By the influence-based user graph, we generate node embeddings with dimensionality $50$ and $100$ for Weibo and Twitter respectively. For semantic embeddings,  in Weibo we used the Chinese word vectors pre-trained on a large Weibo corpus  \cite{li2018analogical}, with word2vec of the Chinese words, while in Twitter we used Glove \cite{pennington2014glove} vectors pre-trained on a large Twitter corpus. The word embedding dimensionality for Twitter is $100$ and for Weibo is $300$. The number of units the attention layer is $200$. In the CNN layer, the number of filters is $120$, and filter length is $3$. The LSTM model has $128$ cells.

\deleted{To avoid biases in our comparison with the baselines, we  kept the same parameter setting for the common stages or components in various models.} \added[id=y]{\emph{We did our best to fine-tune all the baseline methods.}} The knowledge-entity embedding dimensions for DKN and DAN are set to $300$ and $512$ for Weibo \cite{Wang2013XLoreAL} and Twitter \cite{zhu2019graphvite} respectively. For the GERL model, the number of neighbors in user-item and user-user graph is set to $20$, \added[id=y]{while for DICER the number of user-item links is set to $10$ per item, and the number of friends is set to $30$ for users. For the NAML and LSTUR models, the category length is $10$. In NAML-BERT, we use the BERT-Base (12 layers) as the pre-trained model. For the FIM model, we use a $[1, 2, 3]$ hierarchy dilation rate for the convolution layers.} As for DeepWide, the hidden layer structure is $N=[512,512]$, and $N=[256,256]$ for DeepFM and DCN, and the number of factors is $50$ for LibFM. \added[id=y]{Each experiment is repeated 5 times independently, and the averaged results are reported with a $0.95$ confidence.}

\subsection{Results Analysis}
\label{sec:analysis}
We present the comparison with baselines, an analysis on the impact of  hyper-parameters or various components in our model, as well as an analysis of the attention mechanism.
\subsubsection{Comparison with baselines}

\begin{table*}[t!]
\vspace{-6mm}
\caption{Comparison among {\modelname} variants in {\itshape Weibo} \& {\itshape Twitter}}
\vspace{-8mm}
\label{table:baselines}
\begin{center}
\resizebox{\textwidth}{!}{
    \begin{tabular}{lccccccccc}
\hline
\hline
\multirow{2}{*}{\textbf{Models}}&\multicolumn{4}{c}{Weibo}&&\multicolumn{4}{c}{Twitter} \\
\cline{2-5} \cline{7-10}
  & AUC & MRR & NDCG@5 & NDCG@10 && AUC & MRR & NDCG@5 & NDCG@10 \\
\hline
 LibFM & $65.01\pm1.34$  &$13.47\pm0.89$ &$60.79\pm1.10$ & $61.34\pm0.57$ && $65.50\pm1.05$ &$10.25\pm1.09$ &$59.78\pm0.60$ &$61.45\pm0.67$\\
 
DeepFM  &$65.49\pm1.13$&$15.12\pm0.95$
&$61.96\pm1.28$ &$64.57\pm0.98$& &$68.23\pm0.82$&$11.38\pm1.25$
&$63.47\pm0.98$ &$65.76\pm0.98$\\

DeepWide &$66.96\pm0.78$ &$15.79\pm0.81$ &$62.93\pm1.21$ &$65.68\pm0.88$ & & $68.52\pm1.05$&$11.57\pm0.82$ &$63.61\pm0.62$ &$65.88\pm0.59$\\

DCN &$67.12\pm0.96$ &$17.78\pm0.89$ &$64.67\pm0.34$ &$66.34\pm0.67$ & &$68.89\pm0.91$&$11.73\pm0.91$
&$63.97\pm0.76$ &$66.02\pm1.19$\\

\hline
DKN &$68.03\pm1.32$ &$18.01\pm0.85$ &$65.25\pm1.57$ &$66.98\pm1.11$ & &$69.17\pm1.15$&$11.96\pm0.50$
&$65.02\pm0.27$&$67.08\pm1.43$\\

DAN &$68.79\pm0.79$&$18.27\pm0.35$
&$65.45\pm0.77$&$67.49\pm0.82$ & &$69.72\pm0.68$ &$12.05\pm0.91$ &$65.35\pm0.86$ &$67.18\pm1.05$\\

GERL &$69.90\pm1.02 $&$18.42\pm0.63$
&$66.35\pm0.96$&$68.37\pm1.07$ & &$70.21\pm0.38$&$12.32\pm0.34$
&$65.62\pm0.59$&$67.74\pm0.52$\\

NAML &$71.13\pm1.54$ & $19.06\pm0.34$ &$66.76\pm2.08$&$68.23\pm1.47$ & &$70.35\pm0.69$&$12.99\pm0.24$ &$66.83\pm1.22$ &$69.25\pm0.89$\\

NAML-BERT &$71.74\pm1.04$ &$19.44\pm0.26$ &$67.15\pm0.75$ &$69.01\pm1.47$ & &$\underline{71.05\pm0.33}$&$\underline{13.52\pm0.97}$ 
&$\underline{67.20\pm1.45}$ &$\underline{69.80\pm1.52}$\\

LSTUR &$71.02\pm0.87$ &$19.00\pm0.91$ &$66.95\pm0.38$ &$68.57\pm0.65$ & &$70.15\pm0.57$&$12.61\pm1.05$ &$66.54\pm1.77$ &$68.57\pm1.58$\\

FIM &$\underline{72.69\pm0.94}$ &$\underline{19.87\pm0.67}$ 
&$\underline{67.56\pm2.03}$&$\underline{69.52\pm1.74}$ & &$70.68\pm0.89$&$13.02\pm0.81$ &$67.01\pm0.96$ &$69.65\pm0.67$\vspace{0.5mm}\\

\hline
DICER &$69.35\pm0.99$ &$17.92\pm0.68$ &$66.54\pm1.20$ &$68.16\pm0.67$ &  &$70.18\pm0.74$&$12.28\pm0.56$ &$65.63\pm1.22$ &$67.75\pm0.24$\\

\hline
{\modelname}-$A_{t}$ &$73.34\pm0.88$&$21.03\pm0.34$
&$68.77\pm1.45$&$70.56\pm0.57$& &$71.93\pm0.79$&$13.31\pm0.17$
&$69.09\pm0.58$&$71.74\pm0.87$\\

{\modelname}-$V_{t}$ &$72.98\pm1.37$&$20.23\pm0.67$
&$68.06\pm1.13$&$70.07\pm1.75$ & &$71.59\pm1.02$&$13.07\pm0.33$
&$68.76\pm1.02$&$70.93\pm0.67$\\

{\modelname}&$\mathbf{74.57\pm0.93}$&$\mathbf{21.57\pm0.25}$&$\mathbf{69.23\pm0.84}$&$\mathbf{71.11\pm0.79}$ & &$\mathbf{72.71\pm1.19}$&$\mathbf{13.83\pm0.58}$
&$\mathbf{69.75\pm0.85}$&$\mathbf{72.03\pm0.25}$\\

\hline
\hline
\end{tabular}}
\end{center}
\vspace{-5mm}
\end{table*}

 From Table~\ref{table:baselines}, we can see that {\modelname} outperforms all the other methods on all metrics \deleted{(with the sole exception of F1 for DAN)}on news data from social media, which validates our initial motivation. We stress that the users we selected for comparison are those with at least 20 adoptions; the impact on performance of the history size is discussed further in Sec. \ref{sec:attentionv}.

We can see that, generally, LibFM performs worse than other deep learning-based models, which indicates that the factorization model fails to capture high-order connectivity between users and news, nor the interactions among users in the social graph. DeepFM, DeepWide, and DCN are considered a panel of classical CTR prediction models for generalized item recommendation. Within this panel, DeepFM makes up for traditional factorization machines, with its deep part for learning high-dimensional interactions. However, DCN performs best on both datasets, indicating that the cross-network component in the model is effective in dealing with the heterogeneity of information from our scenario.

\eat{
\begin{figure}[t!]
  \centering
  \includegraphics[width=\linewidth]{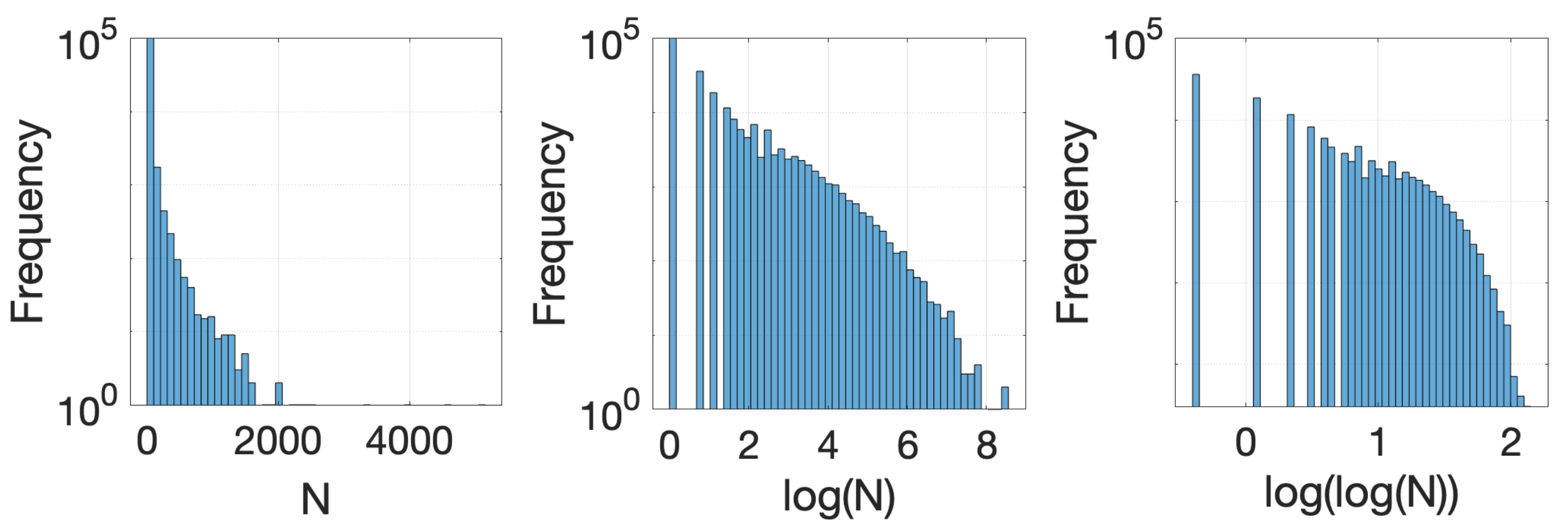}
  \caption{Scaling function for adoption sequences; $N$ is the number of adoption per hour.}
  \label{fig:scaling_fct}
\end{figure}
}

All the neural news recommendation models outperform the general deep recommendation models. This does not come as a surprise, given the particular nature of news and the NLP dimension to be considered. Within this panel of models, \added[id=y]{NAML-BERT and FIM outperform other models in Weibo and Twitter respectively. With a pre-trained language model, NAML-BERT improves the performance of NAML on both datasets, showing the effectiveness of pre-trained language model in news recommendation. FIM proves the importance of fine-grained pairwise multi-level matching between candidate news and a user history (instead of a vector-wise user history view). We can also note that, although DICER is not specifically designed for news recommendation, it performs relatively well, compared to other general recommendation models; this further supports our initial motivation of enhancing news recommendation by the integration of social-aware information.}

 The results of the  ablation study of \modelname \deleted{we also consider two of its main variations, by removing from the news raw representation either the diffusion node sequence $V_t$ or the adoption sequence $A_t$, denoted as {\modelname}-$V_{t}$ and {\modelname}-$A_{t}$ respectively.} show that both the users involved in the  diffusion chain and the adoption pattern play an important role for our recommendation task, with the former contributing slightly more than the latter.
 \subsubsection{Variation with the history length}
\begin{figure}[t!]
\vspace{-1mm}
  \centering
  \includegraphics[width=0.48\linewidth]{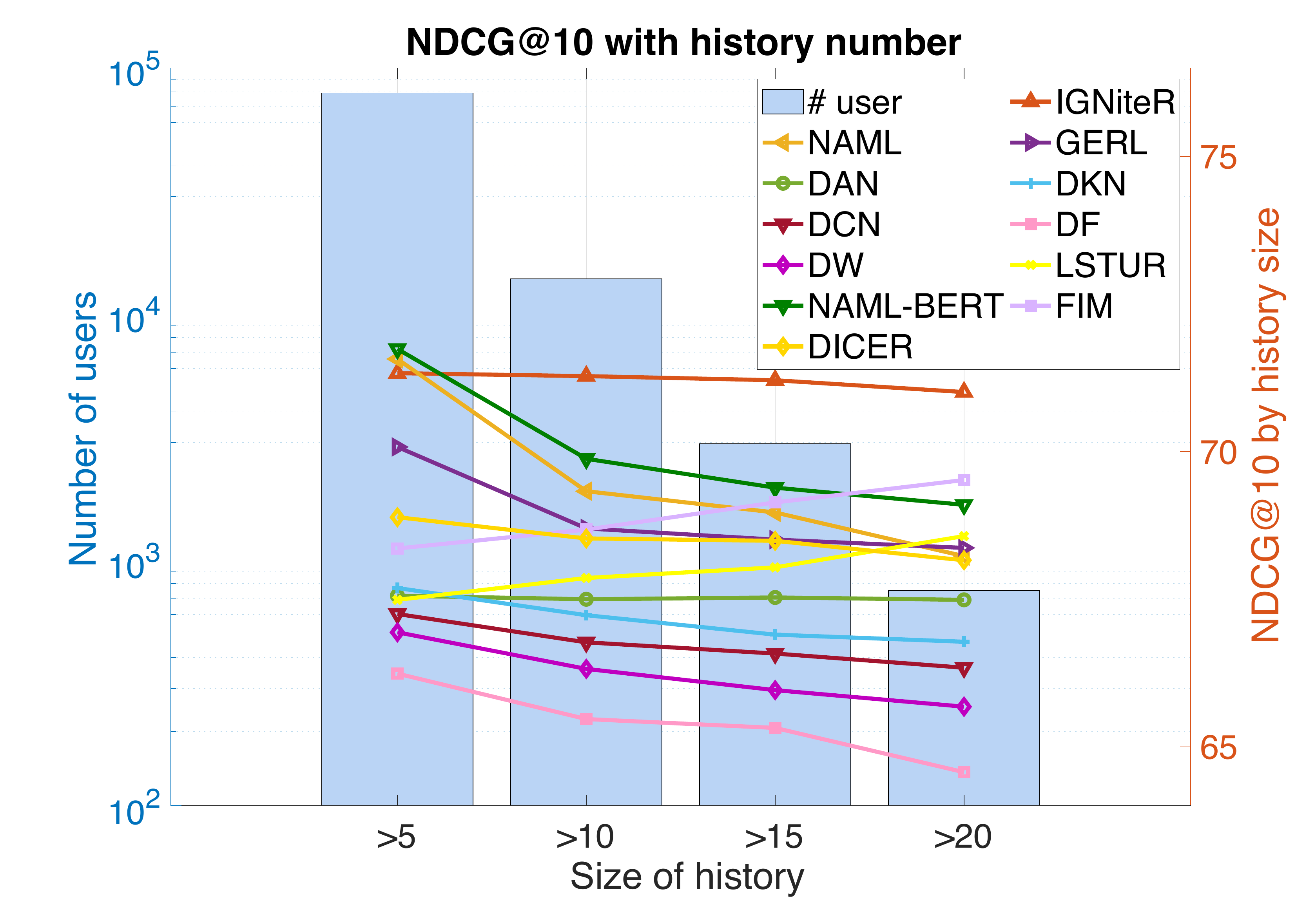}
  \includegraphics[width=0.48\linewidth]{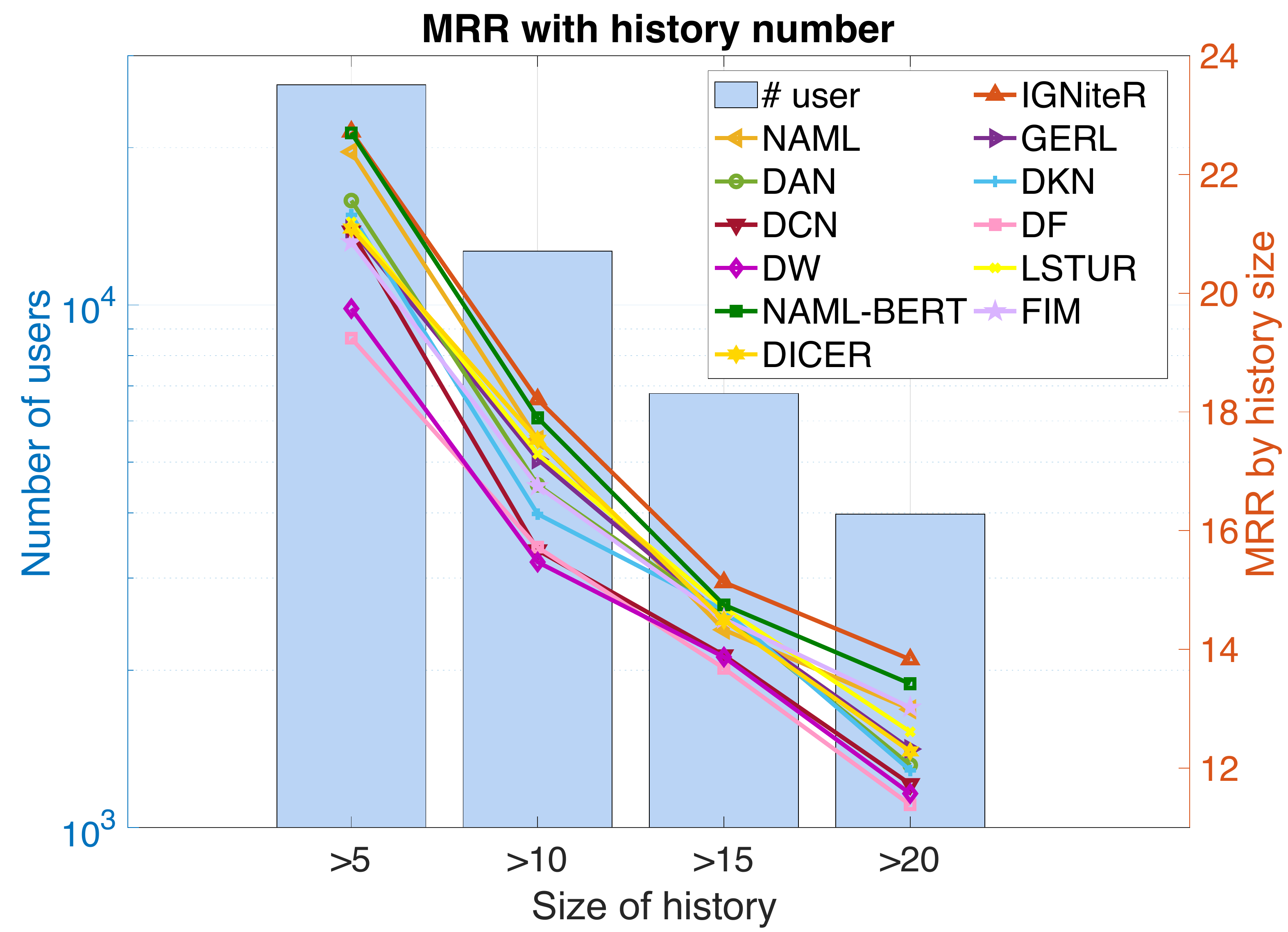}
  \vspace{-2mm}
  \caption{NDCG@10 in Weibo (left) / MRR in Twitter (right) by  history size.}
  \label{fig:ndcg@10}
\vspace{-5mm}
\end{figure}

The size of the users' history in a given time window represents an important factor for recommendation. As it indicates how active and adoption-prone users may be, this can lead to  variations in performance. To exploit this, instead of zero padding or cutting out adoptions, we performed separate experiments for user groups having different activeness levels. We divided the users into four groups according to their overall history (no longer limited to a three months window), as users with more than $20$,  $15$, $10$, and $5$ adopted items. The user count in each group is given in the histograms of  Fig. \ref{fig:ndcg@10}, and  
 we measure NDCG@10 for Weibo and MRR for Twitter. 

Fig. \ref{fig:ndcg@10} (left) shows the variation of performance with the different activeness. \deleted{the performance of all the models decreases.}For the group of user with more than $5$ adoptions, NAML /  NAML-BERT outperform slightly \modelname, \deleted{and GERL does comparably well as \modelname,}while the performance of NAML and GERL decreases drastically for more active users. The overall curve of \modelname\ is slightly decreasing as well, but stabilises at a good level. The performance of FIM and LSTUR goes up as the activeness of users increases. It is interesting to note that DAN has a similar evolution and robustness, even though it generally performs worse than NAML and GERL. \added[id=y]{We can credit this to the LSTM / GRU component present in DAN, LSTUR, and \modelname, capturing dependencies in long sequences. For FIM, our interpretation is that its multi-level matching structure  may be more suitable for picking the salient features when abundant data is available}. Fig. \ref{fig:ndcg@10} (right) shows that with increased history size, the increased informativeness makes the models progressively incapable of placing the positive samples in a conspicuous position.  Hence the performance degrades for all models on the MRR metric; nevertheless, \modelname\ still outperforms the other models in all groups of users.

\begin{figure}[t!]
\vspace{-3mm}
  \centering
  \includegraphics[width=0.7\linewidth]{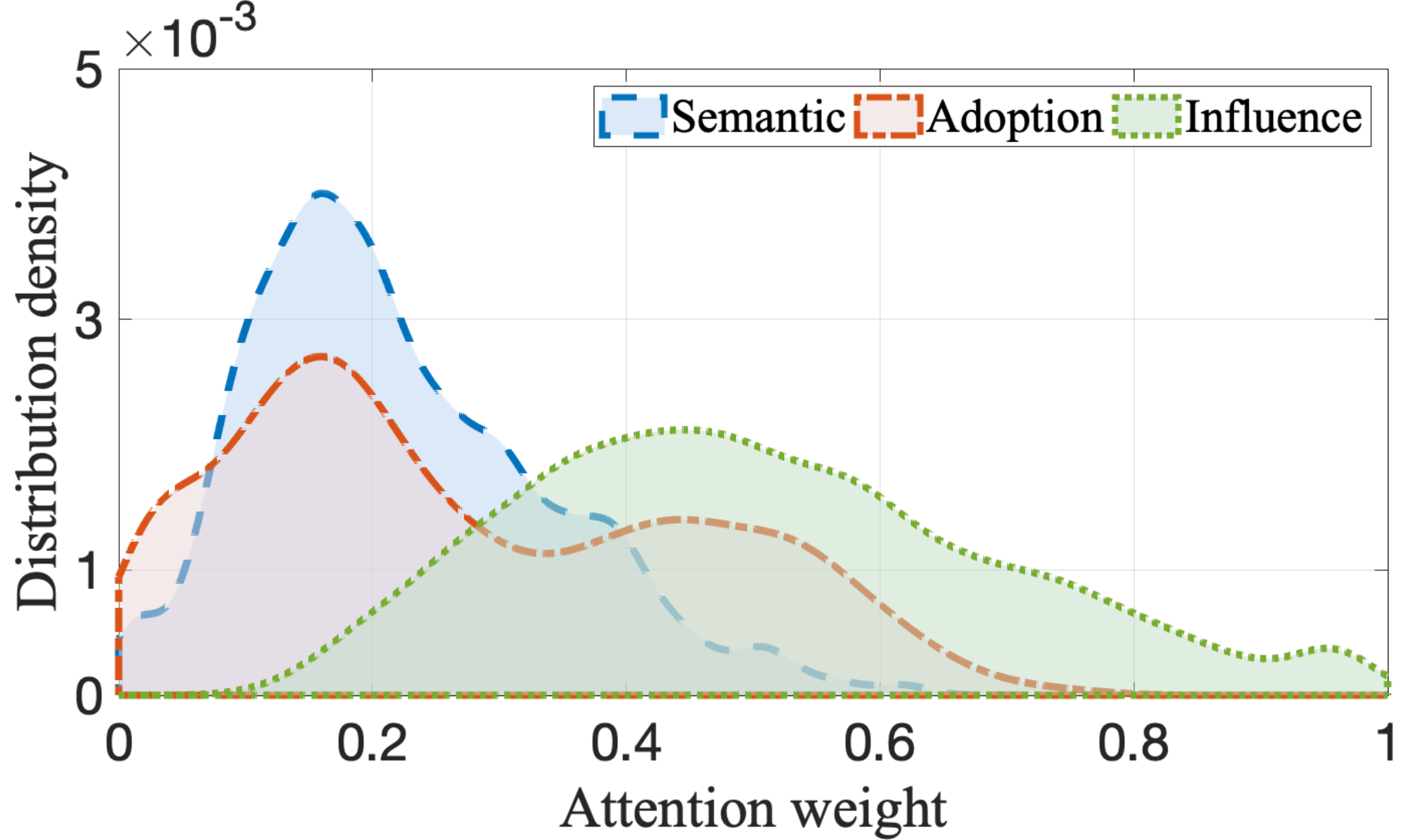}
  \vspace{-2mm}
  \caption{Attention view distribution.}
  \label{fig:att_view}
\vspace{-1mm}
\end{figure}

\begin{figure}[t!]
\vspace{-3mm}
  \centering
  \includegraphics[width=0.8\linewidth]{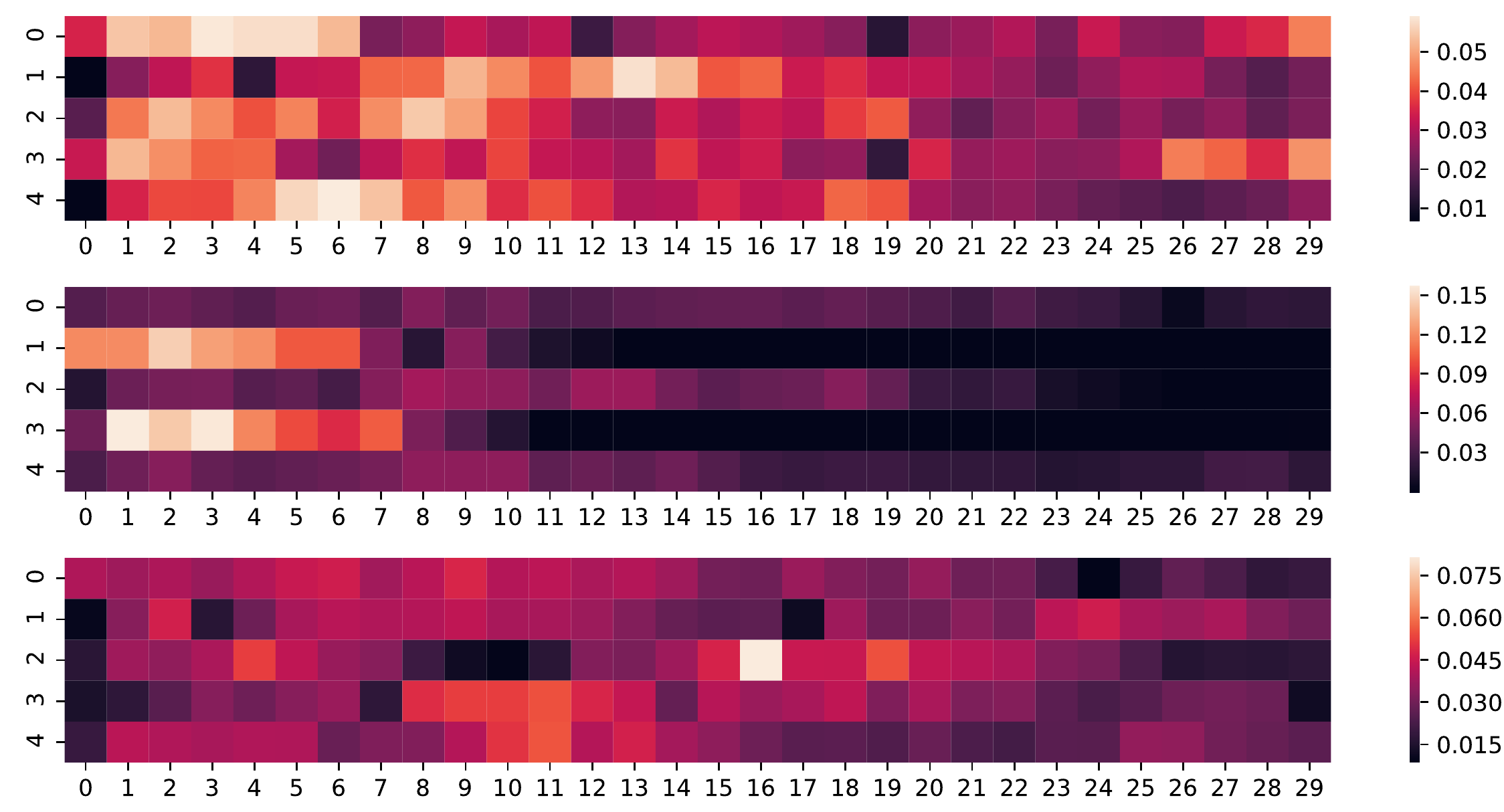}
  \vspace{-2mm}
  \caption{Attention distribution (for each user the heat map is a $5\times30$ matrix, for each news the nodes are noted from 0 to 29, ordered by adoption time).}
  \label{fig:node_attention}
\vspace{-5mm}
\end{figure}

\subsubsection{Attention Visualization}
\label{sec:attentionv}
Complementing the deep neural structure, the attention mechanism enables to expose certain indicators explaining the model's predictions. In this section,  we give some insights on  what can be observed in the case of news recommendations in the microblogging scenario.\vspace{+1mm}\\
\textbf{View attention distribution.} Recall that, with news described initially by $(s^x,V^x_{t},A^x_{t})$, an attention mechanism is applied in the news encoder in Sec.\ref{sec:multi-view attention} to fuse the semantics, diffusion,  and adoption pattern facets. To explore the importance of these dimensions, we randomly sampled $31440$ pieces of news from $15000$ users, in order  to visualize the attention weights allocated to each dimension (Fig. \ref{fig:att_view}). We can see that, generally, the diffusion dimension gets the highest attention weights, which validates once again our motivation. 
 The figure also shows that the adoption pattern dimension -- capturing the timeliness and popularity of news -- gets more attention than the semantics dimension, serving as the main additional signal to refine the recommendation predictions.  
 Interestingly, semantics counts the least in this analysis.  We can conclude from these observations that, in the microblogging scenario, temporal, viral, and influence / credibility aspects are driving the news adoptions decisions.\vspace{+1mm}\\ 
 \textbf{Attention visualization within the diffusion dimension.} 
To give further insights into the diffusion dimension, which weighs the most in making predictions, we sampled $3$ users and charted the distribution of attention weights over nodes in their adoption history. We randomly selected for each user $5$ pieces of news from the  history, for each piece of news we select the $30$ most representative users from the diffusion chain by Alg. \ref{alg:node sampling}, and then fed this into the trained model. The heat map of attention weights is shown in Fig. \ref{fig:node_attention}.
 

The attention weights of the top user are mostly concentrated in the front and middle sections, i.e., the beginning and the middle of propagation chains, which means that the users involved in the early stages of diffusions matter more than those involved later.  
This probably corresponds to a user who tends to follow trendy topics, participating in viral events.  The attention weights for the second user almost all concentrate in the front section, indicating that this is a user who is more likely to be influenced by cascade initiators or by users involved in the early propagation stages,   
so a user who often follows the actions of ``social whales''. Finally, the attention weights of the third user are rather evenly distributed in the propagation chains, with several highlighted users for each piece of news. Unlike the previous two, this user may be influenced by a particular group of users who share common interests, or who are in a close social circle.

\section{Conclusion}
\label{sec:conclusion}

We propose in this paper a content-based deep learning model called {\modelname} for news recommendation, tailored for recommendation scenarios in social media. To incorporate awareness about news due to social influence, we represent users by embeddings obtained by  methods leveraging the diffusion history (cascades), in such a way that news are endowed with diffusion-related information. A CNN method is applied to deal with the joint representation of news and an attention mechanism allows us to aggregate the users' diverse interests with respect to candidate news. Extensive experiments are conducted on two real-world datasets, including a publicly available one, showcasing the significant improvements of {\modelname} over state-of-the-art recommendation models.

\tiny

\balance

\bibliographystyle{IEEEtran}
\bibliography{ref}

\begin{thebibliography}{10}
\providecommand{\url}[1]{#1}
\csname url@samestyle\endcsname
\providecommand{\newblock}{\relax}
\providecommand{\bibinfo}[2]{#2}
\providecommand{\BIBentrySTDinterwordspacing}{\spaceskip=0pt\relax}
\providecommand{\BIBentryALTinterwordstretchfactor}{4}
\providecommand{\BIBentryALTinterwordspacing}{\spaceskip=\fontdimen2\font plus
\BIBentryALTinterwordstretchfactor\fontdimen3\font minus
  \fontdimen4\font\relax}
\providecommand{\BIBforeignlanguage}[2]{{%
\expandafter\ifx\csname l@#1\endcsname\relax
\typeout{** WARNING: IEEEtran.bst: No hyphenation pattern has been}%
\typeout{** loaded for the language `#1'. Using the pattern for}%
\typeout{** the default language instead.}%
\else
\language=\csname l@#1\endcsname
\fi
#2}}
\providecommand{\BIBdecl}{\relax}
\BIBdecl

\bibitem{ricci2014recommender}
F.~Ricci, ``Recommender systems: Models and techniques.'' 2014.

\bibitem{salakhutdinov2007restricted}
R.~Salakhutdinov, A.~Mnih, and G.~Hinton, ``Restricted boltzmann machines for
  collaborative filtering,'' in \emph{ICML}, 2007.

\bibitem{cheng2016wide}
H.-T. Cheng \emph{et~al.}, ``Wide \& deep learning for recommender systems,''
  in \emph{Workshop on deep learning for recommender systems}, 2016.

\bibitem{guo2017deepfm}
H.~Guo, R.~Tang, Y.~Ye, Z.~Li, and X.~He, ``{DeepFM}: a factorization-machine
  based neural network for {CTR} prediction,'' \emph{arXiv:1703.04247}.

\bibitem{wang2018dkn}
H.~Wang, F.~Zhang, X.~Xie, and M.~Guo, ``{DKN}: Deep knowledge-aware network
  for news recommendation,'' in \emph{WWW}, 2018.

\bibitem{zhu2019dan}
Q.~Zhu, X.~Zhou, Z.~Song, J.~Tan, and L.~Guo, ``{DAN}: Deep attention neural
  network for news recommendation,'' in \emph{AAAI}, 2019.

\bibitem{meng2021dcan}
L.~Meng, C.~Shi, S.~Hao, and X.~Su, ``Dcan: Deep co-attention network by
  modeling user preference and news lifecycle for news recommendation,'' in
  \emph{DASFAA}, 2021.

\bibitem{ge2020graph}
S.~Ge, C.~Wu, F.~Wu, T.~Qi, and Y.~Huang, ``Graph enhanced representation
  learning for news recommendation,'' in \emph{WWW}, 2020.

\bibitem{wu2019npa}
C.~Wu, F.~Wu, M.~An, J.~Huang, Y.~Huang, and X.~Xie, ``{NPA}: Neural news
  recommendation with personalized attention,'' in \emph{KDD}, 2019.

\bibitem{fan2019graph}
W.~Fan, Y.~Ma, Q.~Li, Y.~He, E.~Zhao, J.~Tang, and D.~Yin, ``Graph neural
  networks for social recommendation,'' in \emph{WWW}, 2019.

\bibitem{xiao2019beyond}
W.~Xiao, H.~Zhao, H.~Pan, Y.~Song, V.~W. Zheng, and Q.~Yang, ``Beyond
  personalization: Social content recommendation for creator equality and
  consumer satisfaction,'' in \emph{KDD}, 2019.

\bibitem{bughin2015getting}
J.~Bughin, ``Getting a sharper picture of social media’s influence,''
  \emph{McKinsey Quarterly}, 2015.

\bibitem{DBLP:journals/tkdd/Gomez-RodriguezLK12}
M.~Gomez{-}Rodriguez, J.~Leskovec, and A.~Krause, ``Inferring networks of
  diffusion and influence,'' \emph{TKDD}, 2012.

\bibitem{DBLP:journals/toc/KempeKT15}
D.~Kempe, J.~M. Kleinberg, and {\'{E}}.~Tardos, ``Maximizing the spread of
  influence through a social network,'' \emph{Theory Comput.}, 2015.

\bibitem{leskovec2006patterns}
J.~Leskovec, A.~Singh, and J.~Kleinberg, ``Patterns of influence in a
  recommendation network,'' in \emph{PAKDD}, 2006.

\bibitem{SNnews}
\BIBentryALTinterwordspacing
R.~Institute. (2020) {Executive Summary and Key Findings of the 2020 Report}.
  [Online]. Available:
  \url{https://www.digitalnewsreport.org/survey/2020/overview-key-findings-2020}
\BIBentrySTDinterwordspacing

\bibitem{wu2019neural}
C.~Wu, F.~Wu, M.~An, J.~Huang, Y.~Huang, and X.~Xie, ``Neural news
  recommendation with attentive multi-view learning,'' in \emph{IJCAI}, 2019.

\bibitem{wu2019neural2}
C.~Wu, F.~Wu, S.~Ge, T.~Qi, Y.~Huang, and X.~Xie, ``Neural news recommendation
  with multi-head self-attention,'' in \emph{EMNLP-IJCNLP'19}, 2019.

\bibitem{wu2020user}
C.~Wu, F.~Wu, T.~Qi, and Y.~Huang, ``User modeling with click preference and
  reading satisfaction for news recommendation.'' in \emph{IJCAI'20}, 2020.

\bibitem{an2019neural}
M.~An, F.~Wu, C.~Wu, K.~Zhang, Z.~Liu, and X.~Xie, ``Neural news recommendation
  with long-and short-term user representations,'' in \emph{ACL}, 2019.

\bibitem{wang2020fine}
H.~Wang, F.~Wu, Z.~Liu, and X.~Xie, ``Fine-grained interest matching for neural
  news recommendation,'' in \emph{ACL}, 2020.

\bibitem{zhang2021unbert}
Q.~Zhang, J.~Li, Q.~Jia, C.~Wang, J.~Zhu, Z.~Wang, and X.~He, ``Unbert:
  User-news matching bert for news recommendation,'' in \emph{IJCAI}, 2021.

\bibitem{zhang2019deep}
S.~Zhang, L.~Yao, A.~Sun, and Y.~Tay, ``Deep learning based recommender system:
  A survey and new perspectives,'' \emph{ACM CSUR}, 2019.

\bibitem{le2014distributed}
Q.~Le and T.~Mikolov, ``Distributed representations of sentences and
  documents,'' in \emph{ICML}, 2014.

\bibitem{DBLP:conf/naacl/DevlinCLT19}
J.~Devlin, M.~Chang, K.~Lee, and K.~Toutanova, ``{BERT:} pre-training of deep
  bidirectional transformers for language understanding,'' in
  \emph{{NAACL-HLT}}, 2019.

\bibitem{wu2021empowering}
C.~Wu, F.~Wu, T.~Qi, and Y.~Huang, ``Empowering news recommendation with
  pre-trained language models,'' in \emph{SIGIR}, 2021.

\bibitem{okura2017embedding}
S.~Okura, Y.~Tagami, S.~Ono, and A.~Tajima, ``Embedding-based news
  recommendation for millions of users,'' in \emph{KDD}, 2017.

\bibitem{DBLP:journals/corr/abs-2003-00982}
\BIBentryALTinterwordspacing
V.~P. Dwivedi, C.~K. Joshi, T.~Laurent, Y.~Bengio, and X.~Bresson,
  ``Benchmarking graph neural networks,'' \emph{CoRR}, vol. abs/2003.00982,
  2020. [Online]. Available: \url{https://arxiv.org/abs/2003.00982}
\BIBentrySTDinterwordspacing

\bibitem{hu2020graph}
L.~Hu, C.~Li, C.~Shi, C.~Yang, and C.~Shao, ``Graph neural news recommendation
  with long-term and short-term interest modeling,'' \emph{Information
  Processing \& Management}, vol.~57, no.~2, p. 102142, 2020.

\bibitem{wang2019kgat}
X.~Wang, X.~He, Y.~Cao, M.~Liu, and T.-S. Chua, ``Kgat: Knowledge graph
  attention network for recommendation,'' in \emph{KDD}, 2019.

\bibitem{guo2015trustsvd}
G.~Guo, J.~Zhang, and N.~Yorke-Smith, ``Trustsvd: Collaborative filtering with
  both the explicit and implicit influence of user trust and of item ratings,''
  in \emph{AAAI}, 2015.

\bibitem{ma2011recommender}
H.~Ma, D.~Zhou, C.~Liu, M.~R. Lyu, and I.~King, ``Recommender systems with
  social regularization,'' in \emph{WSDM}, 2011.

\bibitem{wu2020diffnet++}
L.~Wu, J.~Li, P.~Sun, R.~Hong, Y.~Ge, and M.~Wang, ``Diffnet++: A neural
  influence and interest diffusion network for social recommendation,''
  \emph{IEEE Transactions on Knowledge and Data Engineering}, 2020.

\bibitem{fu2021dual}
B.~Fu, W.~Zhang, G.~Hu, X.~Dai, S.~Huang, and J.~Chen, ``Dual side deep
  context-aware modulation for social recommendation,'' 2021.

\bibitem{wang2019neural}
X.~Wang, X.~He, M.~Wang, F.~Feng, and T.-S. Chua, ``Neural graph collaborative
  filtering,'' in \emph{SIGIR}, 2019.

\bibitem{panagopoulos2020multi}
G.~Panagopoulos, F.~Malliaros, and M.~Vazirgiannis, ``Multi-task learning for
  influence estimation and maximization,'' 2020.

\bibitem{kim-2014-convolutional}
Y.~Kim, ``Convolutional neural networks for sentence classification,'' in
  \emph{EMNLP}, 2014.

\bibitem{castillo2014characterizing}
C.~Castillo, M.~El-Haddad, J.~Pfeffer, and M.~Stempeck, ``Characterizing the
  life cycle of online news stories using social media reactions,'' in
  \emph{Proceedings of the 17th ACM conference on Computer supported
  cooperative work \& social computing}, 2014, pp. 211--223.

\bibitem{zhai2016deepintent}
S.~Zhai, K.-h. Chang, R.~Zhang, and Z.~M. Zhang, ``Deepintent: Learning
  attentions for online advertising with recurrent neural networks,'' in
  \emph{KDD}, 2016.

\bibitem{zhang2013social}
J.~Zhang, B.~Liu, J.~Tang, T.~Chen, and J.~Li, ``Social influence locality for
  modeling retweeting behaviors,'' in \emph{IJCAI}, 2013.

\bibitem{rendle2012factorization}
S.~Rendle, ``Factorization machines with {LibFM},'' \emph{ACM TIST}, 2012.

\bibitem{wang2020dcn}
R.~Wang, R.~Shivanna, D.~Cheng, S.~Jain, D.~Lin, L.~Hong, and E.~Chi,
  ``{DCN-M}: Improved deep \& cross network for feature cross learning in
  web-scale learning to rank systems,'' \emph{arXiv:2008.13535}.

\bibitem{li2018analogical}
S.~Li, Z.~Zhao, R.~Hu, W.~Li, T.~Liu, and X.~Du, ``Analogical reasoning on
  chinese morphological and semantic relations,'' \emph{arXiv:1805.06504},
  2018.

\bibitem{pennington2014glove}
J.~Pennington, R.~Socher, and C.~D. Manning, ``{Glove}: Global vectors for word
  representation,'' in \emph{EMNLP}, 2014.

\bibitem{Wang2013XLoreAL}
Z.~Wang, J.-Z. Li, Z.~Wang, S.~Li, M.~Li, D.~Zhang, Y.~Shi, Y.~Liu, P.~Zhang,
  and J.~Tang, ``Xlore: A large-scale english-chinese bilingual knowledge
  graph,'' in \emph{SEMWEB}, 2013.

\bibitem{zhu2019graphvite}
Z.~Zhu, S.~Xu, M.~Qu, and J.~Tang, ``Graphvite: A high-performance cpu-gpu
  hybrid system for node embedding,'' in \emph{WWW}, 2019.

\end{thebibliography}

\end{document}